# Longitudinal Mediation Analysis with Latent Growth Curves

Adam J. Sullivan, Douglas D. Gunzler, Nathan Morris, Tyler J. VanderWeele

Abstract. The paper considers mediation analysis with longitudinal data under latent growth curve models within a counterfactual framework. Estimators and their standard errors are derived for natural direct and indirect effects when the mediator, the outcome, and possibly also the exposure can be modeled by an underlying latent variable giving rise to a growth curve. Settings are also considered in which the exposure is instead fixed at a single point in time.

## 1 Introduction

There is a large body of published literature on mediation analysis [1–9]. Almost all of this literature has considered mediation analysis for a single exposure, a single mediator and a single outcome all at one point in time. However in many studies longitudinal data is available and often not used. Instead empirical analysis often rely on the cross sectional models which do not allow for exploiting the temporal sequence of these variables. In addition, it has been shown that cross-sectional mediation analysis typically generates substantially biased estimates of longitudinal parameters even under the ideal conditions when mediation is complete [10]. The use of longitudinal models would allow for less bias and stronger claims of causality.

In the literature there are three main types of longitudinal models currently in use. The models are the autoregressive model [11, 12], latent growth curve models [13–18] and latent difference score models [19–21]. In this paper the focus is on advancing the methodology of mediation with latent growth curve models. We make three major contributions to the literature. We put the models into a formal causal framework so that they may be accurately used to make causal inferences. We then clarify the assumptions needed in order to make causal inferences. Finally we extend existing methodology to allow for interaction to be assessed with these models.

We first consider a latent growth curve model with binary treatment/exposure. With this model we consider the assumptions needed for identifiabilty of the direct and indirect effects. We define the direct and indirect effects, using counterfactuals, in the presence of interaction. We then consider the scenario where there is a longitudinal treatment/exposure. We consider the assumptions needed for identification and define the direct and indirect effects using counterfactuals which allows for the presence of interaction. We finish this paper with an data analysis example.

# 2 Definition of Model

When there is repeated measures data for the mediator and outcome, mediation models can be fit using latent growth curve (LGC) modeling [13–18]. We use the parallel process model as shown by MacKinnon [18] in which separate growth curves are specified for the mediator and outcome. The treatment/exposure can also have a specified growth curve, or as with a randomized trial, it can be binary. With these growth models there are latent factors included. The first of these factors is the intercept or average baseline of the subjects at the first measurement occasion. The second factor is the slope or the trajectory of the growth after the first measurement occasion. When using these models in the mediation setting we examine the mediating relationships of these latent factors among the growth models.

We begin with Model 1 shown in Figure 1. We have a binary treatment, $X_i$; longitudinal mediator, $M_1, M_2$ and $M_3$; and a longitudinal outcome, $Y_1, Y_2$ and $Y_3$. With this model $X_i$ affects both the intercept and slope of the mediator and outcome growth models. The intercept and the slope of the mediator growth model also both effect the intercept and the slope of the outcome growth model.

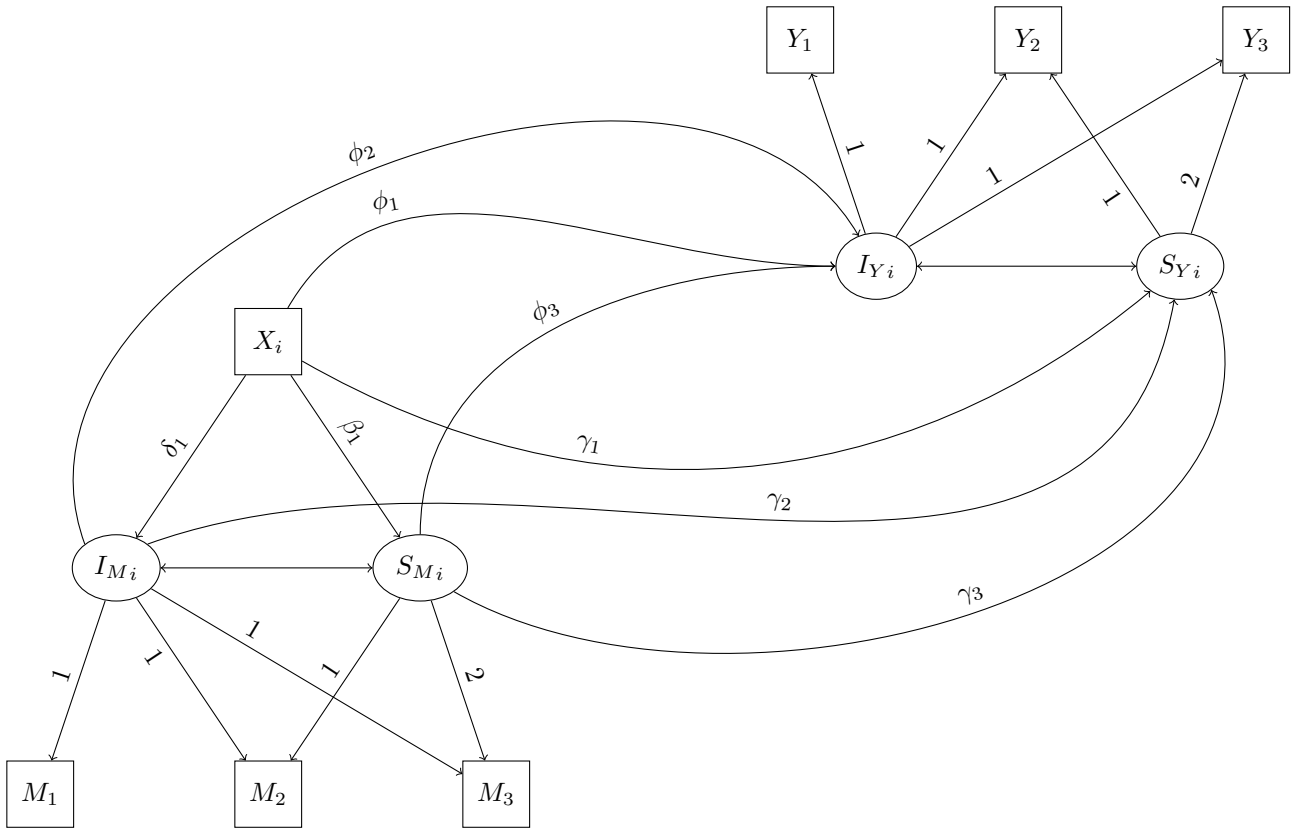

Figure 1: Model 1: Without Interaction, covariates $C$ left out for simplicity

More formally, equations (1) - (6) below specify the relationships shown in Figure 1 with $1, \ldots, t$ measurement occasions. We have the following growth curve for the mediator:

$$M_{it} = I_{Mi} + S_{Mi} t + \varepsilon_{Mit} \tag{1}$$

$$I_{Mi} = \delta_0 + \delta_1 X_i + \delta_2' C + \nu_{I_{Mi}} \tag{2}$$

$$S_{Mi} = \beta_0 + \beta_1 X_i + \beta_2' C + \nu_{S_{Mi}} \tag{3}$$

and the following growth curve for the outcome:

$$Y_{it} = I_{Yi} + S_{YI} t + \varepsilon_{Yit} \tag{4}$$

$$I_{Yi} = \phi_0 + \phi_1 X_i + \phi_2 I_{Mi} + \phi_3 S_{Mi} + \phi_4 X_i I_{Mi} + \phi_5 X_i S_{Mi} + \phi_6' C + \nu_{I_{Yi}} \tag{5}$$

$$S_{Yi} = \gamma_0 + \gamma_1 X_i + \gamma_2 I_{Mi} + \gamma_3 S_{Mi} + \gamma_4 X_i I_{Mi} + \gamma_5 X_i S_{Mi} + \gamma_6' C + \nu_{S_{Yi}} \tag{6}$$

Where $\mathbb{E}\left[\varepsilon_{Mit}\right] = \mathbb{E}\left[\varepsilon_{Yit}\right] = \mathbb{E}\left[\nu_{I_{Mi}}\right] = \mathbb{E}\left[\nu_{S_{Mi}}\right] = \mathbb{E}\left[\nu_{I_{Yi}}\right] = \mathbb{E}\left[\nu_{S_{Yi}}\right] = 0$ and where $\varepsilon_{Mit}, \varepsilon_{Yit}, (\nu_{I_{Mi}}, \nu_{S_{Mi}})$ and $(\nu_{I_{Yi}}, \nu_{S_{Yi}})$ are mutually independent and where $C$ denotes baseline covariates which, as discussed below we select to represent the set of exposure-mediator, exposure-outcome and mediator-outcome confounders.

Equations 1 and 4 specify the growth models for individual $i$'s mediator and outcome data respectively at time $t$. Both include an intercept, slope and error component. Equations 2-3 and 5-6 specify the intercept and slope functions for the mediator and outcome models respectively. Note that equations 5 and 6 allow for treatment/exposure-mediator interaction. In the absence interaction we specify $\phi_4 = \phi_5 = \gamma_4 = \gamma_5 = 0$.

We use counterfactual notation $I_{Yxm_1m_2}$, $S_{Yxm_1m_2}$, $I_{Mx}$ and $S_{Mx}$, where $I_{Yxm_1m_2}$ denotes the value of the intercept model for $Y$ if we were to set $X = x$, $I_M = m_1$ and $S_M = m_2$; $S_{Yxm_1m_2}$ denotes the value of the slope model for $Y$ if we were to set $X = x$, $I_M = m_1$ and $S_M = m_2$; $I_{Mx}$ denotes the value of the intercept model for $M$ if we were to set $X = x$ and $S_{Mx}$ denotes the value of the slope model for $M$ if we were to set $X = x$. We use $Y_{xm_1m_2}$ to denote the counterfactual outcome $Y$ if we were to set $X = x$, $I_M = m_1$ and $S_M = m_2$. The natural direct effect for two values of the exposure, $x$ and $x^*$, is defined as $\mathbb{E}\left[Y_{xI_{Mx^*}S_{Mx^*}} - Y_{x^*I_{Mx^*}S_{Mx^*}}\right]$ and expresses how much the intercept and slope of the outcome process would change on average if the treatment/exposure were changed from level $x^* = 0$ to $x = 1$ but for each individual the intercept and slope of the mediator process is kept at the level it would have taken under the absence of the treatment/exposure. The natural indirect effect for two values of the exposure, $x$ and $x^*$, is defined as $\mathbb{E}\left[Y_{xI_{M_x}S_{M_x}} - Y_{xI_{M_{x^*}}S_{M_{x^*}}}\right]$ and expresses how much the intercept and slope of the outcome process would change on average if the treatment/exposure was controlled at level $x = 1$ but the intercept and slope of the mediator process were changed from the level they would take if the treatment/exposure was changed from $x^* = 0$ to $x = 1$. We let $A \amalg B | C$ denote that $A$ is independent of $B$ conditional on $C$. We show below that the natural direct and indirect effects are identified if:

$$I_{YI_MS_M}, S_{YI_MS_M} \amalg X | C \quad \text{(no unmeasured confounding for the exposure-outcome relationship)} \quad \text{(C1)}$$

$$I_{YI_MS_M}, S_{YI_MS_M} \amalg I_M, S_M | X, C \quad \text{(no unmeasured confounding for the mediator-outcome relationship)} \quad \text{(C2)}$$

$$I_{M_x}, S_{M_x} \amalg X | C \quad \text{(no unmeasured confounding for the exposure-mediator relationship)} \quad \text{(C3)}$$

$$I_{Ym_1,m_2}, S_{Ym_1,m_2} \amalg I_{M_{x^*}}, S_{M_{x^*}} | C \quad \text{(no mediator-outcome confounders which are affected by the exposure)} \quad \text{(C4)}$$

Proposition: For any function $u$ if (C1) - (C4) hold then

$$\mathbb{E}\left[u\left(I_{YxI_{M_{x^*}}S_{M_{x^*}}}, S_{YxI_{M_{x^*}}S_{M_{x^*}}}\right)\right] = \sum_{c,m_1,m_2} \mathbb{E}\left[u\left(I_Y, S_Y\right) | x, m_1, m_2, c\right] \Pr\left(m_1, m_2 | x^*, c\right) \Pr(c)$$

Proof:

$$\mathbb{E}\left[u\left(I_{Y x I_{M x^*} S_{M x^*}}, S_{Y x I_{M x^*} S_{M x^*}}\right)\right] = \sum_c \mathbb{E}\left[u\left(I_{Y x I_{M x^*} S_{M x^*}}, S_{Y x I_{M x^*} S_{M x^*}}\right) | C = c\right] \Pr(C = c) \quad \text{(Iterated Expectations)}$$

$$= \sum_{c, m_1, m_2} \mathbb{E}\left[u\left(I_{Y x m_1 m_2}, S_{Y x m_1 m_2}\right) | C = c, I_{M x^*} = m_1, S_{M x^*} = m_2\right] \Pr\left(I_{M x^*} = m_1, S_{M x^*} = m_2 | C = c\right) \Pr(C = c)$$

(Iterated Expectations)

$$= \sum_{c, m_1, m_2} \mathbb{E}\left[u\left(I_{Y x m_1 m_2}, S_{Y x m_1 m_2}\right) | C = c\right] \Pr\left(I_{M x^*} = m_1, S_{M x^*} = m_2 | X = x^*, C = c\right) \Pr(C = c) \quad \text{(C4 \& C3)}$$

$$= \sum_{c, m_1, m_2} \mathbb{E}\left[u\left(I_{Y x m_1 m_2}, S_{Y x m_1 m_2}\right) | X = x, C = c\right] \Pr\left(I_M = m_1, S_M = m_2 | X = x^*, C = c\right) \Pr(C = c) \quad \text{(C1 \& consistency)}$$

$$= \sum_{c, m_1, m_2} \mathbb{E}\left[u\left(I_{Y x m_1 m_2}, S_{Y x m_1 m_2}\right) | X = x, I_M = m_1, S_M = m_2, C = c\right] \Pr\left(I_M = m_1, S_M = m_2 | X = x^*, C = c\right) \Pr(C = c)$$

(C2)

$$= \sum_{c, m_1, m_2} \mathbb{E}\left[u\left(I_Y, S_Y\right) | X = x, I_M = m_1, S_M = m_2, C = c\right] \Pr\left(I_M = m_1, S_M = m_2 | X = x^*, C = c\right) \Pr(C = c)$$

(consistency)

$$= \sum_{c, m_1, m_2} \mathbb{E}\left[u\left(I_Y, S_Y\right) | x, m_1, m_2, c\right] \Pr\left(m_1, m_2 | x^*, c\right) \Pr(c)$$

This completes the proof.

Then if we replace $x$ with $x^*$ we get:

$$\mathbb{E}\left[u\left(I_{Y x^* I_{M x^*} S_{M x^*}}, S_{Y x^* I_{M x^*} S_{M x^*}}\right)\right] = \sum_{c, m_1, m_2} \mathbb{E}\left[S_Y | x^*, m_1, m_2, c\right] \Pr\left(m_1, m_2 | x^*, c\right) \Pr(c)$$

from this it follows with $u(I_Y, S_Y) = I_Y + S_Y t + \varepsilon_Y$ that the average natural direct effect is given by:

$$\mathbb{E}\left[u\left(I_{Y x I_{M x^*} S_{M x^*}}, S_{Y x I_{M x^*} S_{M x^*}}\right) - u\left(I_{Y x^* I_{M x^*} S_{M x^*}}, S_{Y x^* I_{M x^*} S_{M x^*}}\right)\right]$$
$$= \sum_{c, m_1, m_2} \left\{\mathbb{E}\left[u\left(I_Y, S_Y\right) | x, m_1, m_2, c\right] - \mathbb{E}\left[u\left(I_Y, S_Y\right) | x^*, m_1, m_2, c\right]\right\} \Pr\left(m_1, m_2 | x^*, c\right) \Pr(c)$$

If we replace $x^*$ with $x$ we would get:

$$\mathbb{E}\left[u\left(I_{Y x I_{M x} S_{M x}}, S_{Y x I_{M x} S_{M x}}\right)\right] = \sum_{c, m_1, m_2} \mathbb{E}\left[u) I_Y, S_Y) | x, m_1, m_2, c\right] \Pr\left(m_1, m_2 | x, c\right) \Pr(c)$$

from this it follows with $u(I_Y, S_Y) = I_Y + S_Y t + \varepsilon_Y$ that the natural indirect effect is given by:

$$\mathbb{E}\left[u\left(I_{YxI_{M_x}S_{M_x}}, S_{YxI_{M_x}S_{M_x}}\right) - u\left(I_{YxI_{M_{x^*}}S_{M_{x^*}}}, S_{YxI_{M_{x^*}}S_{M_{x^*}}}\right)\right]$$

$$= \sum_{c,m_1,m_2} \mathbb{E}\left[u\left(I_Y, S_Y\right) | x, m_1, m_2, c\right] \left\{\Pr(m_1, m_2 | x, c) - \Pr(m_1, m_2 | x^*, c)\right\} \Pr(c)$$

$$= \sum_{c,m_1,m_2} \mathbb{E}\left[u\left(I_Y, S_Y\right) | x, m_1, m_2, c\right] \Pr(m_1, m_2 | x, c) \Pr(c) - \mathbb{E}\left[u\left(I_Y, S_Y\right) | x, m_1, m_2, c\right] \Pr(m_1, m_2 | x^*, c) \Pr(c)$$

With the model shown in Figure 1 we have that $Y = u(I_Y, S_Y) = I_Y + S_Y t + \epsilon_Y$. Thus given (5) and (6) we have

$$\mathbb{E}\left[u\left(I_Y, S_Y\right) | x, m_1, m_2, c\right] = \phi_0 + \phi_1 x + \phi_2 m_1 + \phi_3 m_2 + \phi_4 x m_1 + \phi_5 x m_2 + \phi_6' c$$
$$+ (\gamma_0 + \gamma_1 x + \gamma_2 m_1 + \gamma_3 m_2 + \gamma_4 x m_1 + \gamma_5 x m_2 + \gamma_6' c) t \quad (7)$$

and

$$\mathbb{E}\left[u\left(I_Y, S_Y\right) | x^*, m_1, m_2, c\right] = \phi_0 + \phi_1 x^* + \phi_2 m_1 + \phi_3 m_2 + \phi_4 x^* m_1 + \phi_5 x^* m_2 + \phi_6' c$$
$$+ (\gamma_0 + \gamma_1 x^* + \gamma_2 m_1 + \gamma_3 m_2 + \gamma_4 x^* m_1 + \gamma_5 x^* m_2 + \gamma_6' c) t \quad (8)$$

Therefore the average natural direct effect is

$$\sum_{c,m_1,m_2} \left\{(\phi_1 + \phi_4 m_1 + \phi_5 m_2 + \gamma_1 t + \gamma_4 m_1 t + \gamma_5 m_2 t)(x - x^*)\right\} \Pr(m_1, m_2 | x^*, c) \Pr(c)$$

$$= (\phi_1 + \phi_4 \mathbb{E}\left[M_1 | x^*, c\right] + \phi_5 \mathbb{E}\left[M_2 | x^*, c\right] + \gamma_1 t + \gamma_4 \mathbb{E}\left[M_1 | x^*, c\right] t + \gamma_5 \mathbb{E}\left[M_2 | x^*, c\right] t)(x - x^*)$$

$$= (\phi_1 + \phi_4(\delta_0 + \delta_1 x^* + \delta_3' c) + \phi_5(\beta_0 + \beta_1 x^* + \beta_3' c) + \gamma_1 t + \gamma_4(\delta_0 + \delta_1 x^* + \delta_2' c) t + \gamma_5(\beta_0 + \beta_1 x^* + \beta_2' c) t)(x - x^*)$$

Given (2), (3), (5) and (6) we have

$$\mathbb{E}\left[u\left(I_{YxI_{M_x}S_{M_x}}, S_{YxI_{M_x}S_{M_x}}\right)\right] = (\phi_2 + \gamma_2 t)\delta_0 + (\phi_3 + \gamma_3 t)\beta_0 + \phi_0 + \gamma_0 t$$
$$+ (\phi_1 + \gamma_1 t + (\phi_2 + \gamma_2 t)\delta_1 + (\phi_3 + \gamma_3 t)\beta_1 + (\phi_4 + \gamma_4 t)(\delta_0 + \delta_2' c) + (\phi_5 + \gamma_5 t)(\beta_0 + \beta_2' c))x$$
$$+ ((\phi_4 + \gamma_4 t)\delta_1 + (\phi_5 + \gamma_5 t)\beta_1)x^2 + (\phi_6' + \gamma_6' t + (\phi_2 + \gamma_2 t)\delta_2' + (\phi_3 + \gamma_3 t)\beta_2')c$$

$$\mathbb{E}\left[u\left(I_{YxI_{Mx^*}S_{Mx^*}}, S_{YxI_{Mx^*}S_{Mx^*}}\right)\right] = (\phi_2 + \gamma_2 t)\delta_0 + (\phi_3 + \gamma_3 t)\beta_0 + \phi_0 + \gamma_0 t$$
$$+ (\phi_1 + \gamma_1 t + (\phi_4 + \gamma_4 t)(\delta_0 + \delta_2' c) + (\phi_5 + \gamma_5 t)(\beta_0 + \beta_2' c))x + ((\phi_2 + \gamma_2 t)\delta_1 + (\phi_3 + \gamma_3 t)\beta_1)x^*$$
$$+ ((\phi_4 + \gamma_4 t)\delta_1 + (\phi_5 + \gamma_5 t)\beta_1)xx^* + (\phi_6' + \gamma_6' t + (\phi_2 + \gamma_2 t)\delta_2' + (\phi_3 + \gamma_3 t)\beta_2')c$$

Therefore the average natural indirect effect is

$$\sum_{c,m_1,m_2} \mathbb{E}\left[u\left(I_Y, S_Y\right) | x, m_1, m_2, c\right] \Pr(m_1, m_2 | x, c) \Pr(c) - \mathbb{E}\left[u\left(I_Y, S_Y\right) | x, m_1, m_2, c\right] \Pr(m_1, m_2 | x^*, c) \Pr(c)$$
$$= (\phi_1 + \gamma_1 t + (\phi_2 + \gamma_2 t)\delta_1 + (\phi_3 + \gamma_3 t)\beta_1 + (\phi_4 + \gamma_4 t)(\delta_0 + \delta_2' c) + (\phi_5 + \gamma_5 t)(\beta_0 + \beta_2' c))x$$
$$+ ((\phi_4 + \gamma_4 t)\delta_1 + (\phi_5 + \gamma_5 t)\beta_1)x^2 - ((\phi_4 + \gamma_4 t)\delta_1 + (\phi_5 + \gamma_5 t)\beta_1)xx^*$$
$$- (\phi_1 + \gamma_1 t + (\phi_4 + \gamma_4 t)(\delta_0 + \delta_2' c) + (\phi_5 + \gamma_5 t)(\beta_0 + \beta_2' c))x - ((\phi_2 + \gamma_2 t)\delta_1 + (\phi_3 + \gamma_3 t)\beta_1)x^*$$
$$= ((\phi_2 + \gamma_2 t)\delta_1 + (\phi_3 + \gamma_3 t)\beta_1)(x - x^*) + ((\phi_4 + \gamma_4 t)\delta_1 + (\phi_5 + \gamma_5 t)\beta_1)(x^2 - xx^*)$$

As discussed previously in the absence of interaction we specify $\phi_4 = \phi_5 = \gamma_4 = \gamma_5 = 0$. This leads to the following direct effect:

$$\sum_{c,m_1,m_2} \{(\phi_1 + \phi_4 m_1 + \phi_5 m_2 + \gamma_1 t + \gamma_4 m_1 t + \gamma_5 m_2 t)(x - x^*)\} \Pr(m_1, m_2 | x^*, c) \Pr(c)$$
$$= (\phi_1 + \gamma_1 t)(x - x^*)$$

and the following indirect effect:

$$\sum_{c,m_1,m_2} \mathbb{E}\left[u\left(I_Y, S_Y\right) | x, m_1, m_2, c\right] \Pr(m_1, m_2 | x, c) \Pr(c) - \mathbb{E}\left[u\left(I_Y, S_Y\right) | x, m_1, m_2, c\right] \Pr(m_1, m_2 | x^*, c) \Pr(c)$$
$$= ((\phi_2 + \gamma_2 t)\delta_1 + (\phi_3 + \gamma_3 t)\beta_1)(x - x^*)$$

The equations modeled here differ from that of the ones presented by MacKinnon [18] in that the intercept of the outcome is not a cause of the slope of the mediator. This follows because if $I_Y$ affected $S_M$ then confounding assumption (C4) would be violated because $I_Y$ would be a mediator-outcome confounder (i.e. a common cause of $S_M$ and $Y$) that was itself affected by the exposure. These equations here unlike those of MacKinnon also allow for exposure/treatment-mediator interaction.

# 3 Model with Growth Curve for Treatment/Exposure

In Section 2 the models were developed under the assumption of a binary treatment/exposure. This is often the case in randomized trials. In this section we consider the model displayed in Figure 2. This model allows for the treatment/exposure to change with time and fits a growth curve for this as well.

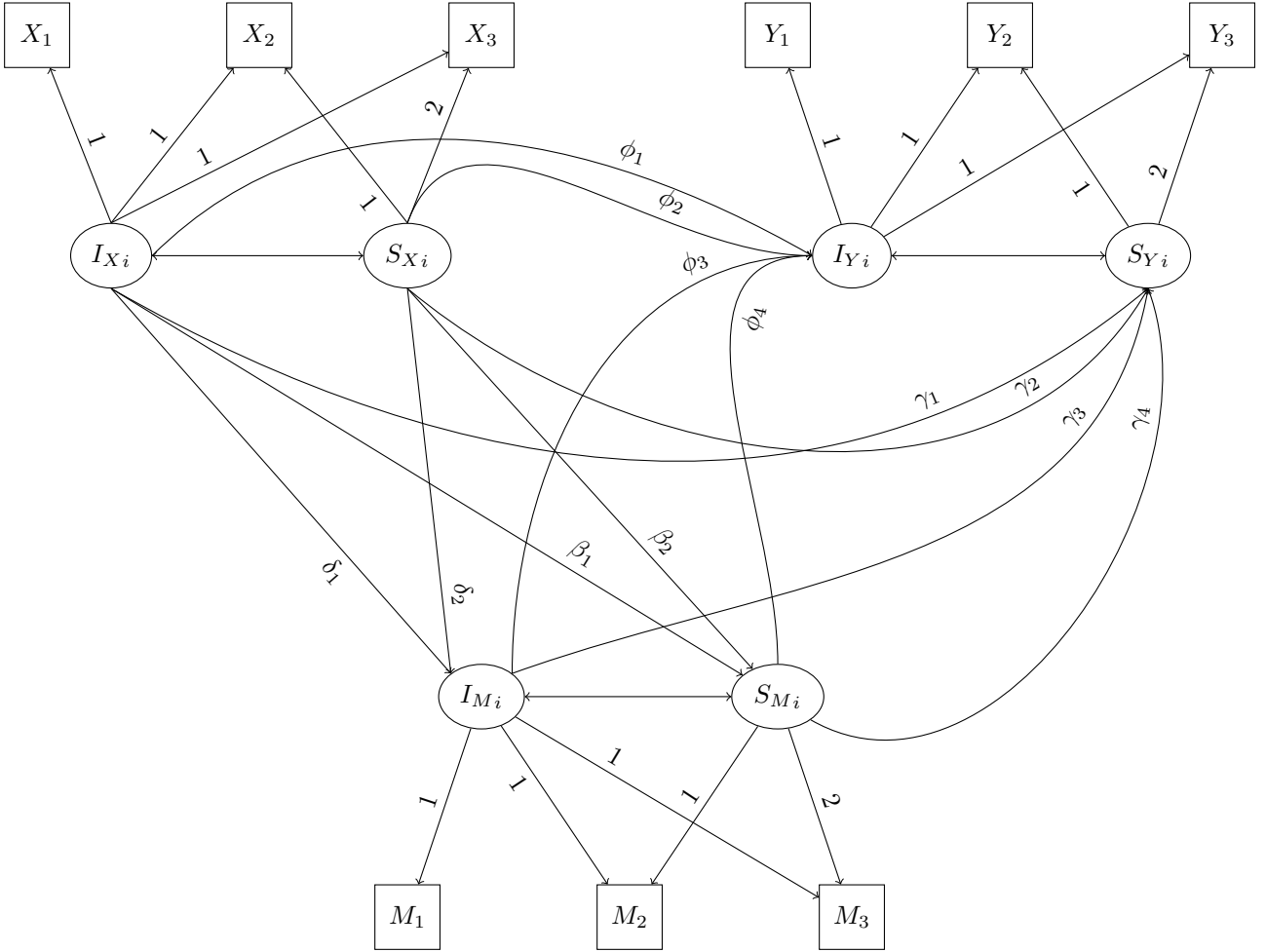

Figure 2: Model 2: Without Interaction, covariates $C$ left out for simplicity

More formally, equations (9) - (17) specify the relationships shown in Figure 2 with $1,\ldots,t$ measurement occasions. We have the following growth curve for the treatment/exposure:

$$X_{it} = I_{Xi} + S_{Xi}t + \varepsilon_{Xit} \tag{9}$$

$$I_{Xi} = \rho_0 + \nu_{I_{Xi}} \tag{10}$$

$$S_{Xi} = \lambda_0 + \nu_{S_{Xi}} \tag{11}$$

the following growth curve for the mediator:

$$M_{it} = I_{Mi} + S_{Mi}t + \varepsilon_{Mit} \tag{12}$$

$$I_{Mi} = \delta_0 + \delta_1 I_{Xi} + \delta_2 S_{Xi} + \delta_3' C + \nu_{I_{Mi}} \tag{13}$$

$$S_{Mi} = \beta_0 + \beta_1 I_{Xi} + \beta_2 S_{Xi} + \beta_3' C + \nu_{I_{Mi}} \tag{14}$$

and the following growth curve for the outcome:

$$Y_{it} = I_{Yi} + S_{YI} t + \varepsilon_{Yit} \tag{15}$$

$$I_{Yi} = \phi_0 + \phi_1 I_{Xi} + \phi_2 S_{Xi} + \phi_3 I_{Mi} + \phi_4 S_{Mi} + \phi_5 I_{Xi} I_{Mi} + \phi_6 I_{Xi} S_{Mi} + \phi_7 S_{Xi} I_{Mi} + \phi_8 S_{Xi} S_{Mi} + \phi_9' C + \nu_{I_{Yi}} \tag{16}$$

$$S_{Yi} = \gamma_0 + \gamma_1 I_{Xi} + \gamma_2 S_{Xi} + \gamma_3 I_{Mi} + \gamma_4 S_{Mi} + \gamma_5 I_{Xi} I_{Mi} + \gamma_6 I_{Xi} S_{Mi} + \gamma_7 S_{Xi} I_{Mi} + \gamma_8 S_{Xi} S_{Mi} + \gamma_9' C + \nu_{S_{Yi}} \tag{17}$$

Where $\mathbb{E}[\varepsilon_{Xit}] = \mathbb{E}[\varepsilon_{Mit}] = \mathbb{E}[\varepsilon_{Yit}] = \mathbb{E}[\nu_{I_{Xi}}] = \mathbb{E}[\nu_{S_{Xi}}] = \mathbb{E}[\nu_{I_{Mi}}] = \mathbb{E}[\nu_{S_{Mi}}] = \mathbb{E}[\nu_{I_{Yi}}] = \mathbb{E}[\nu_{S_{Yi}}] = 0$ and where $\varepsilon_{Xit}, \varepsilon_{Mit}, \varepsilon_{Yit}, (\nu_{I_{Xi}}, \nu_{S_{Xi}}), (\nu_{I_{Mi}}, \nu_{S_{Mi}})$ and $(\nu_{I_{Yi}}, \nu_{S_{Yi}})$ are mutually independent and where $C$ denotes baseline covariates which as discussed below we select to represent exposure-mediator, exposure-outcome and mediator-outcome confounders.

Equations 9, 12 and 15 specify the growth model for individual $i$'s treatment/exposure, mediator and outcome respectively. Equations 10-11, 13-14 and 16-17 specify the intercept and slope for the exposure, mediator and outcome respectively. Note that equations 16 and 17 allow for exposure-mediator interaction. In the absence of interaction we specify $\phi_5 = \phi_6 = \phi_7 = \phi_8 = \gamma_5 = \gamma_6 = \gamma_7 = \gamma_8 = 0$.

We use counterfactual notation $I_{Yx_1x_2m_1m_2}$, $S_{Yx_1x_2m_1m_2}$, $I_{Mx_1x_2}$ and $S_{Mx_1x_2}$, where $I_{Yx_1x_2m_1m_2}$ denotes the value of the intercept model for $Y$ if we were to set $I_X = x_1, S_X = x_2, I_M = m_1$ and $S_M = m_2$; $S_{Yx_1x_2m_1m_2}$ denotes the value of the slope model for $Y$ if we were to set $I_X = x_1, S_X = x_2, I_M = m_1$ and $S_M = m_2$; $I_{Mx_1x_2}$ denotes the value of the intercept model for $M$ if we were to set $I_X = x_1$ and $S_X = x_2$ and $S_{Mx_1x_2}$ denotes the value of the slope model for $M$ if we were to set $X = x$. We use $Y_{x_1x_2m_1m_2}$ to denote the counterfactual outcome $Y$ if we were to set $I_X = x_1, S_X = x_2, I_M = m_1$ and $S_M = m_2$. The natural direct effect for two values of the intercept function of the treatment/exposure $x_1$ and $x_1^*$ and for two values of the slope function of the treatment/exposure $x_2$ and $x_2^*$, is defined as $\mathbb{E}\left[Y_{x_1x_2 I_{Mx_1^*x_2^*} S_{Mx_1^*x_2^*}} - Y_{x_1^*x_2^* I_{Mx_1^*x_2^*} S_{Mx_1^*x_2^*}}\right]$ and expresses how much the intercept and slope of the outcome process would change on average if the intercept and slope functions of the treatment/exposure were changed from levels $x_1^* = x_2^* = 0$ to $x_1 = a_1$ and $x_2 = a_2$ but for each individual the intercept and slope of the mediator process is kept at the level it would have taken under the absence of the treatment/exposure. The natural indirect effect for two values of the intercept function of the treatment/exposure $x_1$ and $x_1^*$ and for two values of the slope function of the treatment/exposure $x_2$ and $x_2^*$, is defined as $\mathbb{E}\left[Y_{x_1x_2 I_{Mx_1x_2} S_{Mx_1x_2}} - Y_{x_1x_2 I_{Mx_1^*x_2^*} S_{Mx_1^*x_2^*}}\right]$ and expresses how much the intercept and slope of the outcome process would change on average if the if the intercept and slope functions of the treatment/exposure were controlled at levels $x_1 = a_1$ and $x_2 = a_2$ but the intercept and slope of the mediator process were changed from the level they would take if the if the intercept and slope functions of the treatment/exposure functions were changed from $x_1^* = x_2^* = 0$ to $x_1 = a_1$ and $x_2 = a_2$. We show below that the natural direct and indirect effects are identified if:

$$I_{Y_{I_M S_M}}, S_{Y_{I_M S_M}} \amalg I_X, S_X | C \quad \text{(no unmeasured confounding for the exposure-outcome relationship)} \quad \text{(C5)}$$

$$I_{Y_{I_M S_M}}, S_{Y_{I_M S_M}} \amalg I_M, S_M | I_X, S_X, C \quad \text{(no unmeasured confounding for the mediator-outcome relationship)} \quad \text{(C6)}$$

$$I_{M_{x_1 x_2}}, S_{M_{x_1 x_2}} \amalg I_X, S_X | C \quad \text{(no unmeasured confounding for the exposure-mediator relationship)} \quad \text{(C7)}$$

$$I_{Y_{m_1, m_2}}, S_{Y_{m_1, m_2}} \amalg I_{M_{x_1^* x_2^*}}, S_{M_{x_1^* x_2^*}} | C \quad \text{(no mediator-outcome confounders which are affected by the exposure)} \quad \text{(C8)}$$

Proposition: For any function $u$ if (C5) - (C-8) hold then

$$\mathbb{E}\left[u\left(I_{Y_{x_1 x_2 I_{M_{x_1^* x_2^*}} S_{M_{x_1^* x_2^*}}}}, S_{Y_{x_1 x_2 I_{M_{x_1^* x_2^*}} S_{M_{x_1^* x_2^*}}}}\right)\right] = \sum_{c, m_1, m_2} \mathbb{E}\left[u(I_Y, S_Y) | x_1, x_2, m_1, m_2, c\right] \Pr(m_1, m_2 | x_1^*, x_2^*, c) \Pr(c)$$

Proof:

$$\mathbb{E}\left[u\left(I_{Y x_1 x_2 I_{M x_1^* x_2^*} S_{M x_1^* x_2^*}}, S_{Y x_1 x_2 I_{M x_1^* x_2^*} S_{M x_1^* x_2^*}}\right)\right]$$

$$= \sum_c \mathbb{E}\left[u\left(I_{Y x_1 x_2 I_{M x_1^* x_2^*} S_{M x_1^* x_2^*}}, S_{Y x_1 x_2 I_{M x_1^* x_2^*} S_{M x_1^* x_2^*}}\right) | C = c\right] \Pr(C=c) \quad \text{(Iterated Expectations)}$$

$$= \sum_{c,m_1,m_2} \mathbb{E}\left[u\left(I_{Y x_1 x_2 m_1 m_2}, S_{Y x_1 x_2 m_1 m_2}\right) | C = c, I_{M x_1^* x_2^*} = m_1, S_{M x_1^* x_2^*} = m_2\right]$$
$$\times \Pr\left(I_{M x_1^* x_2^*} = m_1, S_{M x_1^* x_2^*} = m_2 | C = c\right) \Pr(C = c) \quad \text{(Iterated Expectations)}$$

$$= \sum_{c,m_1,m_2} \mathbb{E}\left[u\left(I_{Y x_1 x_2 m_1 m_2}, S_{Y x_1 x_2 m_1 m_2}\right) | C = c\right] \Pr\left(I_{M x_1^* x_2^*} = m_1, S_{M x_1^* x_2^*} = m_2 | I_X = x_1^*, S_X = x_2^*, C = c\right) \Pr(C = c)$$
$$\text{((C8) \& (C7))}$$

$$= \sum_{c,m_1,m_2} \mathbb{E}\left[u\left(I_{Y x_1 x_2 m_1 m_2}, S_{Y x_1 x_2 m_1 m_2}\right) | I_X = x_1, S_X = x_2, C = c\right]$$
$$\times \Pr\left(I_M = m_1, S_M = m_2 | I_X = x_1^*, S_X = x_2^*, C = c\right) \Pr(C = c) \quad \text{((C5) \& consistency)}$$

$$= \sum_{c,m_1,m_2} \mathbb{E}\left[u\left(I_{Y x_1 x_2 m_1 m_2}, S_{Y x_1 x_2 m_1 m_2}\right) | I_X = x_1, S_X = x_2, I_M = m_1, S_M = m_2, C = c\right]$$
$$\times \Pr\left(I_M = m_1, S_M = m_2 | I_X = x_1^*, S_X = x_2^*, C = c\right) \Pr(C = c) \quad \text{((C6))}$$

$$= \sum_{c,m_1,m_2} \mathbb{E}\left[u\left(I_Y, S_Y\right) | I_X = x_1, S_X = x_2, I_M = m_1, S_M = m_2, C = c\right]$$
$$\times \Pr\left(I_M = m_1, S_M = m_2 | I_X = x_1^*, S_X = x_2^*, C = c\right) \Pr(C = c) \quad \text{(consistency)}$$

$$= \sum_{c,m_1,m_2} \mathbb{E}\left[u\left(I_Y, S_Y\right) | x_1, x_2, m_1, m_2, c\right] \Pr\left(m_1, m_2 | x_1^*, x_2^*, c\right) \Pr(c)$$

This completes the proof.

Then if we replace $x$ with $x^*$ we get:

$$\mathbb{E}\left[u\left(I_{Y x_1^* x_2^* I_{M x_1^* x_2^*} S_{M x_1^* x_2^*}}, S_{Y x_1^* x_2^* I_{M x_1^* x_2^*} S_{M x_1^* x_2^*}}\right)\right] = \sum_{c,m_1,m_2} \mathbb{E}\left[u\left(I_Y, S_Y\right) | x_1^*, x_2^*, m_1, m_2, c\right] \Pr\left(m_1, m_2 | x_1^*, x_2^*, c\right) \Pr(c)$$

from this it follows with $u(I_Y, S_Y) = I_Y + S_Y t + \varepsilon_Y$ that the average natural direct effect is given by:

$$\mathbb{E}\left[u\left(I_{Y x_1 x_2 I_{M x_1^* x_2^*} S_{M x_1^* x_2^*}}, S_{Y x_1 x_2 I_{M x_1^* x_2^*} S_{M x_1^* x_2^*}}\right) - u\left(I_{Y x_1^* x_2^* I_{M x_1^* x_2^*} S_{M x_1^* x_2^*}}, S_{Y x_1^* x_2^* I_{M x_1^* x_2^*} S_{M x_1^* x_2^*}}\right)\right] =$$
$$\sum_{c,m_1,m_2} \left\{\mathbb{E}\left[u\left(I_Y, S_Y\right) | x_1, x_2, m_1, m_2, c\right] - \mathbb{E}\left[u\left(I_Y, S_Y\right) | x_1^*, x_2^*, m_1, m_2, c\right]\right\} \Pr\left(m_1, m_2 | x_1^*, x_2^*, c\right) \Pr(c) \quad (18)$$

If we replaced $x^*$ with $x$ we would get:

$$\mathbb{E}\left[u\left(I_{Y x_1 x_2 I_{M x_1 x_2} S_{M x_1 x_2}}, S_{Y x_1 x_2 I_{M x_1 x_2} S_{M x_1 x_2}}\right)\right] = \sum_{c,m_1,m_2} \mathbb{E}\left[u\left(I_Y, S_Y\right) | x_1, x_2, m_1, m_2, c\right] \Pr\left(m_1, m_2 | x_1, x_2, c\right) \Pr(c)$$

from this it follows with $u(I_Y, S_Y) = I_Y + S_Y t + \varepsilon_Y$ that the natural indirect effect is given by:

$$\mathbb{E}\left[u\left(I_{Y x_1 x_2 I_{M x_1 x_2} S_{M x_1 x_2}}, S_{Y x_1 x_2 I_{M x_1 x_2} S_{M x_1 x_2}}\right) - u\left(I_{Y x_1 x_2 I_{M x_1^* x_2^*} S_{M x_1^* x_2^*}}, S_{Y x_1 x_2 I_{M x_1^* x_2^*} S_{M x^*}}\right)\right] =$$

$$\sum_{c, m_1, m_2} \mathbb{E}\left[u(I_Y, S_Y) | x_1, x_2, m_1, m_2, c\right] \{\Pr(m_1, m_2 | x_1, x_2, c) - \Pr(m_1, m_2 | x_1^*, x_2^*, c)\} \Pr(c) \quad (19)$$

With the model shown in Figure 2 we have that $Y = u(I_Y, S_Y) = I_Y + S_Y t + \epsilon_Y$. Thus given (16) and (17) we have

$$\mathbb{E}\left[u(I_Y, S_Y) | x_1, x_2, m_1, m_2, c\right] = \phi_0 + \phi_1 x_1 + \phi_2 x_2 + \phi_3 m_1 + \phi_4 m_2 + \phi_5 x_1 m_1 + \phi_6 x_1 m_2 + \phi_7 x_2 m_1 + \phi_8 x_2 m_2 + \phi_9' c$$
$$+ (\gamma_0 + \gamma_1 x_1 + \gamma_2 x_2 + \gamma_3 m_1 + \gamma_4 m_2 + \gamma_5 x_1 m_1 + \gamma_6 x_1 m_2 + \gamma_7 + x_2 m_1 + \gamma_8 x_2 m_2 + \gamma_9' c) t$$

and

$$\mathbb{E}\left[u(I_Y, S_Y) | x_1^*, x_2^*, m_1, m_2, c\right] = \phi_0 + \phi_1 x_1^* + \phi_2 x_2^* + \phi_3 m_1 + \phi_4 m_2 + \phi_5 x_1^* m_1 + \phi_6 x_1^* m_2 + \phi_7 x_2^* m_1 + \phi_8 x_2^* m_2 + \phi_9' c$$
$$+ (\gamma_0 + \gamma_1 x_1^* + \gamma_2 x_2^* + \gamma_3 m_1 + \gamma_4 m_2 + \gamma_5 x_1^* m_1 + \gamma_6 x_1^* m_2 + \gamma_7 + x_2^* m_1 + \gamma_8 x_2^* m_2 + \gamma_9' c) t$$

Therefore the average natural direct effect is

$$\sum_{c, m_1, m_2} \{(\phi_1 + \gamma_1 t + (\phi_5 + \gamma_5 t) m_1 + (\phi_6 + \gamma_6 t) m_2)(x_1 - x_1^*) + (\phi_2 + \gamma_2 t + (\phi_7 + \gamma_7 t) m_1 + (\phi_8 + \gamma_8 t) m_2)(x_2 - x_2^*)\}$$
$$= (\phi_1 + \gamma_1 t + (\phi_5 + \gamma_5 t) \mathbb{E}[M_1 | x_1^*, x_2^*, c] + (\phi_6 + \gamma_6 t) \mathbb{E}[M_2 | x_1^*, x_2^*, c])(x_1 - x_1^*) + (\phi_2 + \gamma_2 t + (\phi_7 + \gamma_7 t) \mathbb{E}[M_1 | x_1^*, x_2^*, c]$$
$$+ (\phi_8 + \gamma_8 t) \mathbb{E}[M_2 | x_1^*, x_2^*, c])(x_2 - x_2^*)$$
$$= (\phi_1 + \gamma_1 t + (\phi_5 + \gamma_5 t)(\delta_0 + \delta_1 x_1^* + \delta_2 x_2^* + \delta_3' c) + (\phi_6 + \gamma_6 t)(\beta_0 + \beta_1 x_1^* + \beta_2 x_2^* + \beta_3' c))(x_1 - x_1^*)$$
$$+ (\phi_2 + \gamma_2 t + (\phi_7 + \gamma_7 t)(\delta_0 + \delta_1 x_1^* + \delta_2 x_2^* + \delta_3' c) + (\phi_8 + \gamma_8 t)(\beta_0 + \beta_1 x_1^* + \beta_2 x_2^* + \beta_3' c))(x_2 - x_2^*)$$

Given (13), (14), (15) and (16) we have

$$\mathbb{E}\left[u\left(I_{Y x_1 x_2 I_{M x_1 x_2} S_{M x_1 x_2}}, S_{Y x_1 x_2 I_{M x_1 x_2} S_{M x_1 x_2}}\right)\right] = \phi_0 + \gamma_0 t + (\phi_1 + \gamma_1 t) x_1 + (\phi_2 + \gamma_2 t) x_2$$
$$+ (\phi_3 + \gamma_3 t)(\delta_0 + \delta_3' c) + (\phi_4 + \gamma_4 t)(\beta_0 + \beta_3' c) + ((\phi_5 + \gamma_5 t)(\delta_0 + \delta_3' c) + (\phi_6 + \gamma_6 t)(\beta_0 + \beta_3' c)) x_1$$
$$+ ((\phi_7 + \gamma_7 t)(\delta_0 + \delta_3' c) + (\phi_8 + \gamma_8 t)(\beta_0 + \beta_3' c)) x_2 + ((\phi_3 + \gamma_3 t)\delta_1 + (\phi_4 + \gamma_4 t)\beta_1) x_1$$
$$+ ((\phi_3 + \gamma_3 t)\delta_2 + (\phi_4 + \gamma_4 t)\beta_2) x_2 + ((\phi_5 + \gamma_5 t)\delta_1 + (\phi_6 + \gamma_6 t)\beta_1) x_1 x_1$$
$$+ ((\phi_5 + \gamma_5 t)\delta_2 + (\phi_6 + \gamma_6 t)\beta_2) x_1 x_2 + ((\phi_7 + \gamma_7 t)\delta_1 + (\phi_8 + \gamma_8 t)\beta_1) x_2 x_1$$
$$+ ((\phi_7 + \gamma_7 t)\delta_2 + (\phi_8 + \gamma_8 t)\beta_2) x_2 x_2$$

$$\mathbb{E}\left[u\left(I_{Yx_1x_2 I_{M_{x_1x_2}} S_{M_{x_1x_2}}}, S_{Yx_1x_2 I_{M_{x_1x_2}} S_{M_{x_1x_2}}}\right)\right] = \phi_0 + \gamma_0 t + (\phi_1 + \gamma_1 t)x_1 + (\phi_2 + \gamma_2 t)x_2$$
$$+ (\phi_3 + \gamma_3 t)(\delta_0 + \delta_3' c) + (\phi_4 + \gamma_4 t)(\beta_0 + \beta_3' c) + ((\phi_5 + \gamma_5 t)(\delta_0 + \delta_3' c) + (\phi_6 + \gamma_6 t)(\beta_0 + \beta_3' c))x_1$$
$$+ ((\phi_7 + \gamma_7 t)(\delta_0 + \delta_3' c) + (\phi_8 + \gamma_8 t)(\beta_0 + \beta_3' c))x_2 + ((\phi_3 + \gamma_3 t)\delta_1 + (\phi_4 + \gamma_4 t)\beta_1)x_1^*$$
$$+ ((\phi_3 + \gamma_3 t)\delta_2 + (\phi_4 + \gamma_4 t)\beta_2)x_2^* + ((\phi_5 + \gamma_5 t)\delta_1 + (\phi_6 + \gamma_6 t)\beta_1)x_1 x_1^*$$
$$+ ((\phi_5 + \gamma_5 t)\delta_2 + (\phi_6 + \gamma_6 t)\beta_2)x_1 x_2^* + ((\phi_7 + \gamma_7 t)\delta_1 + (\phi_8 + \gamma_8 t)\beta_1)x_2 x_1^*$$
$$+ ((\phi_7 + \gamma_7 t)\delta_2 + (\phi_8 + \gamma_8 t)\beta_2)x_2 x_2^*$$

Therefore the average natural indirect effect is:

$$\sum_{c,m_1,m_2} \mathbb{E}\left[u\left(I_Y, S_Y\right)|x_1,x_2,m_1,m_2,c\right]\Pr(m_1,m_2|x_1,x_2,c)\Pr(c) - \mathbb{E}\left[u\left(I_Y,S_Y\right)|x_1,x_2,m_1,m_2,c\right]\Pr(m_1,m_2|x_1^*,x_2^*,c)\Pr(c)$$
$$= ((\phi_3 + \gamma_3 t)\delta_1 + (\phi_4 + \gamma_4 t)\beta_1)x_1 + ((\phi_3 + \gamma_3 t)\delta_2 + (\phi_4 + \gamma_4 t)\beta_2)x_2 + ((\phi_5 + \gamma_5 t)\delta_1 + (\phi_6 + \gamma_6 t)\beta_1)x_1 x_1$$
$$+ ((\phi_5 + \gamma_5 t)\delta_2 + (\phi_6 + \gamma_6 t)\beta_2)x_1 x_2 + ((\phi_7 + \gamma_7 t)\delta_1 + (\phi_8 + \gamma_8 t)\beta_1)x_2 x_1 + ((\phi_7 + \gamma_7 t)\delta_2 + (\phi_8 + \gamma_8 t)\beta_2)x_2 x_2$$
$$- ((\phi_3 + \gamma_3 t)\delta_1 + (\phi_4 + \gamma_4 t)\beta_1)x_1^* - ((\phi_3 + \gamma_3 t)\delta_2 + (\phi_4 + \gamma_4 t)\beta_2)x_2^* - ((\phi_5 + \gamma_5 t)\delta_1 + (\phi_6 + \gamma_6 t)\beta_1)x_1 x_1^*$$
$$- ((\phi_5 + \gamma_5 t)\delta_2 + (\phi_6 + \gamma_6 t)\beta_2)x_1 x_2^* - ((\phi_7 + \gamma_7 t)\delta_1 + (\phi_8 + \gamma_8 t)\beta_1)x_2 x_1^* - ((\phi_7 + \gamma_7 t)\delta_2 + (\phi_8 + \gamma_8 t)\beta_2)x_2 x_2^*$$
$$= ((\phi_3 + \gamma_3 t)\delta_1 + (\phi_4 + \gamma_4 t)\beta_1)(x_1 - x_1^*) + ((\phi_3 + \gamma_3 t)\delta_2 + (\phi_4 + \gamma_4 t)\beta_2)(x_2 - x_2^*)$$
$$+ ((\phi_5 + \gamma_5 t)\delta_1 + (\phi_6 + \gamma_6 t)\beta_1)(x_1 x_1 - x_1 x_1^*) + ((\phi_5 + \gamma_5 t)\delta_2 + (\phi_6 + \gamma_6 t)\beta_2)(x_1 x_2 - x_1 x_2^*)$$
$$+ ((\phi_7 + \gamma_7 t)\delta_1 + (\phi_8 + \gamma_8 t)\beta_1)(x_2 x_1 - x_2 x_1^*) + ((\phi_7 + \gamma_7 t)\delta_2 + (\phi_8 + \gamma_8 t)\beta_2)(x_2 x_2 - x_2 x_2^*)$$

As discussed previously in the absence of interaction we specify $\phi_5 = \phi_6 = \phi_7 = \phi_8 = \gamma_5 = \gamma_6 = \gamma_7 = \gamma_8 = 0$. This leads to the following direct effect:

$$\sum_{c,m_1,m_2} \{(\phi_1 + \gamma_1 t)(x_1 - x_1^*) + (\phi_2 + \gamma_2 t)(x_2 - x_2^*)\}\Pr(m_1,m_2|x^*,c)\Pr(c)$$
$$= \{(\phi_1 + \gamma_1 t)(x_1 - x_1^*) + (\phi_2 + \gamma_2 t)(x_2 - x_2^*)\}\sum_{c,m_1,m_2}\Pr(m_1,m_2|x^*,c)\Pr(c)$$
$$= (\phi_1 + \gamma_1 t)(x_1 - x_1^*) + (\phi_2 + \gamma_2 t)(x_2 - x_2^*)$$

and the following indirect effect:

$$\sum_{c,m_1,m_2} \mathbb{E}\left[u\left(I_Y, S_Y\right) | x_1, x_2, m_1, m_2, c\right] \Pr(m_1, m_2 | x_1, x_2, c) \Pr(c) - \mathbb{E}\left[u\left(I_Y, S_Y\right) | x_1, x_2, m_1, m_2, c\right] \Pr(m_1, m_2 | x_1^*, x_2^*, c) \Pr(c)$$

$$= ((\phi_3 + \gamma_3 t)\delta_1 + (\phi_4 + \gamma_4 t)\beta_1)(x_1 - x_1^*) + ((\phi_3 + \gamma_3 t)\delta_2 + (\phi_4 + \gamma_4 t)\beta_2)(x_2 - x_2^*)$$

## 4 Standard Errors of Direct and Indirect Effects

When considering the direct and indirect effects it is important to be able to test the statistical significance of these effects. Folmer[22], Sobel[23, 24], Bollen[25] and Bollen & Stine[26] suggest applying the delta method to estimate the asymptotic variances of the indirect and total effect. We suggest that the delta method is used in this case as well. In the latent growth mediation context both the direct and indirect effects are nonlinear functions of several model coefficient estimators. We then use the first order multivariate delta method in order to approximate the standard errors:

$$g\left(\hat{\theta}\right) \approx g(\theta) + \frac{\partial g(\theta)}{\partial \theta} \qquad (20)$$

Considering equation 20 we see that $g(\hat{\theta})$ is approximately equal to a linear function of $\theta$. We have from large sample theory that $g\left(\hat{\theta}\right)$ is approximately normal. Given that $g(\theta)$ is a constant we have a constant plus a multiple of a normally distributed variable so in large samples $g\left(\hat{\theta}\right)$ is approximately normal[26].

$$g\left(\theta'\right) \sim N\left(g(\theta), \nabla g(\theta)' Var(\theta) \nabla g(\theta)\right)$$

This means that we can use the normal distribution to create confidence intervals as well as perform hypothesis tests on the direct and indirect effects of models 1 and 2.

### 4.1 Standard Errors for Model 1

Using standard SEM software to fit model 1 results in estimates $\hat{\delta}$ of $\delta \equiv (\delta_0, \delta_1, \delta_2')'$, $\hat{\beta}$ of $\beta \equiv (\beta_0, \beta_1, \beta_2')'$, $\hat{\phi}$ of $\phi \equiv (\phi_0, \phi_1, \phi_2, \phi_3, \phi_4, \phi_5, \phi_6')'$ and $\hat{\gamma}$ of $\gamma \equiv (\gamma_0, \gamma_1, \gamma_2, \gamma_3, \gamma_4, \gamma_5, \gamma_6')'$. Using these we take

$$\theta \equiv (\delta, \beta, \phi, \gamma) \equiv (\delta_0, \delta_1, \delta_2', \beta_0, \beta_1, \beta_2', \phi_0, \phi_1, \phi_2, \phi_3, \phi_4, \phi_5, \phi_6', \gamma_0, \gamma_1, \gamma_2, \gamma_3, \gamma_4, \gamma_5, \gamma_6').$$

Given the direct effect for model 1 in Section 2 we have

$$g(\theta) = (\phi_1 + \phi_4(\delta_0 + \delta_1 x^* + \delta_2' c) + \phi_5(\beta_0 + \beta_1 x^* + \beta_2' c) + \gamma_1 t + \gamma_4(\delta_0 + \delta_1 x^* + \delta_2' c)t + \gamma_5(\beta_0 + \beta_1 x^* + \beta_2' c)t$$

Thus we have

$$\nabla g(\theta) = \big(\phi_4 + \gamma_4 t, (\phi_4 + \gamma_4 t)x^*, (\phi_4 + \gamma_4 t)c', \phi_5 + \gamma_5 t, (\phi_5 + \gamma_5 t)x^*, (\phi_5 + \gamma_5 t)c', 0, 1$$
$$, 0, 0, \delta_0 + \delta_1 x^* + \delta_2' c, \beta_0 + \beta_1 x^* + \beta_2' c, 0', 0, 0, t, 0, 0, (\delta_0 + \delta_1 x^* + \delta_2' c)t, (\beta_0 + \beta_1 x^* + \beta_2' c)t, 0' \big)'$$

Thus $SE(g(\theta)) = \sqrt{\nabla g(\theta)' Var(\theta) \nabla g(\theta)}$. This leads to the standard error of the direct effect in model 1:

$$\sqrt{\nabla g(\theta)' Var(\theta) \nabla g(\theta)} |x - x^*|$$

Given the indirect effect for model 1 in Section 2 we have

$$g(\theta) = [(\phi_2 + \gamma_2 t)\delta_1 + (\phi_3 + \gamma_3 t)\beta_1](x - x^*) + [(\phi_4 + \gamma_4 t)\delta_1 + (\phi_5 + \gamma_5 t)\beta_1(x^2 - xx^*)]$$

Thus we have

$$\nabla g(\theta) = \big(0, (\phi_2 + \gamma_2 t)(x - x^*) + (\phi_4 + \gamma_4 t)(x^2 - xx^*), 0', 0, (\phi_3 + \gamma_3 t)(x - x^*) + (\phi_3 + \gamma_5 t)(x^2 - xx^*), 0', 0, 0$$
$$, \delta_1(x - x^*), \beta_1(x - x^*), \delta_1(x^2 - xx^*), \beta_1(x^2 - xx^*), 0', 0, \delta_1 t(x - x^*), \beta_1 t(x - x^*), \delta_1 t(x^2 - xx^*), \beta_1 t(x^2 - xx^*), 0' \big)$$

Thus the standard error of the indirect effect in model 1:

$$\sqrt{\nabla g(\theta)' Var(\theta) \nabla g(\theta)}$$

## 4.2 Standard Errors for Model 2

Using standard SEM software to fit model 2 results in estimates $\hat{\rho}_0$ of $\rho_0$, $\hat{\lambda}_0$ of $\lambda_0$, $\hat{\delta}$ of $\delta \equiv (\delta_0, \delta_1, \delta_2, \delta_3')'$, $\hat{\beta}$ of $\beta \equiv (\beta_0, \beta_1, \beta_2, \beta_3')'$, $\hat{\phi}$ of $\phi \equiv (\phi_0, \phi_1, \phi_2, \phi_3, \phi_4, \phi_5, \phi_6, \phi_7, \phi_8, \phi_9')'$ and $\hat{\gamma}$ of $\gamma \equiv (\gamma_0, \gamma_1, \gamma_2, \gamma_3, \gamma_4, \gamma_5, \gamma_6, \gamma_7, \gamma_8, \gamma_9')'$. Using these we take

$$\theta \equiv (\rho_0, \lambda_0, \delta, \beta, \phi, \gamma) \equiv (\rho_0, \lambda_0, \delta_0, \delta_1, \delta_2, \delta_3', \beta_0, \beta_1, \beta_2, \beta_3', \phi_0, \phi_1, \phi_2, \phi_3, \phi_4, \phi_5, \phi_6, \phi_7, \phi_8, \phi_9', \gamma_0, \gamma_1, \gamma_2, \gamma_3, \gamma_4, \gamma_5, \gamma_6, \gamma_7, \gamma_8, \gamma_9').$$

Given the direct effect for model 2 in Section 3 we have

$$g(\theta) = (\phi_1 + \gamma_1 t + (\phi_5 + \gamma_5 t)(\delta_0 + \delta_1 x_1^* + \delta_2 x_2^* + \delta_3' c) + (\phi_6 + \gamma_6 t)(\beta_0 + \beta_1 x_1^* + \beta_2 x_2^* + \beta_3' c))(x_1 - x_1^*)$$
$$+ (\phi_2 + \gamma_2 t + (\phi_7 + \gamma_7 t)(\delta_0 + \delta_1 x_1^* + \delta_2 x_2^* + \delta_3' c) + (\phi_8 + \gamma_8 t)(\beta_0 + \beta_1 x_1^* + \beta_2 x_2^* + \beta_3' c)(x_2 - x_2^*)$$

Thus we have

$$\nabla g(\theta) = (0, 0, (\phi_5 + \gamma_5 t)(x_1 - x_1^*) + (\phi_7 + \gamma_7 t)(x_2 - x_2^*), (\phi_5 + \gamma_5 t)x_1^*(x_1 - x_1^*) + (\phi_7 + \gamma_7 t)x_1^*(x_2 - x_2^*)$$
$$, (\phi_5 + \gamma_5 t)x_2^*(x_1 - x_1^*) + (\phi_7 + \gamma_7 t)x_2^*(x_2 - x_2^*), (\phi_5 + \gamma_5 t)c'(x_1 - x_1^*) + (\phi_7 + \gamma_7 t)c'(x_2 - x_2^*)$$
$$, (\phi_6 + \gamma_6 t)(x_1 - x_1^*) + (\phi_8 + \gamma_8 t)(x_2 - x_2^*), (\phi_6 + \gamma_6 t)x_1^*(x_1 - x_1^*) + (\phi_8 + \gamma_8 t)x_1^*(x_2 - x_2^*)$$
$$, (\phi_6 + \gamma_6 t)x_2^*(x_1 - x_1^*) + (\phi_8 + \gamma_8 t)x_2^*(x_2 - x_2^*), (\phi_6 + \gamma_6 t)c'(x_1 - x_1^*) + (\phi_8 + \gamma_8 t)c'(x_2 - x_2^*)$$
$$, 0, (x_1 - x_1^*), (x_2 - x_2^*), 0, 0, (\delta_0 + \delta_1 x_1^* + \delta_2 x_2^* + \delta_3'c)(x_1 - x_1^*), (\beta_0 + \beta_1 x_1^* + \beta_2 x_2^* + \beta_3'c)(x_1 - x_1^*)$$
$$, (\delta_0 + \delta_1 x_1^* + \delta_2 x_2^* + \delta_3'c)(x_2 - x_2^*), (\beta_0 + \beta_1 x_1^* + \beta_2 x_2^* + \beta_3'c)(x_2 - x_2^*), 0', 0, (x_1 - x_1^*)t, (x_2 - x_2^*)t$$
$$, 0, 0, (\delta_0 + \delta_1 x_1^* + \delta_2 x_2^* + \delta_3'c)(x_1 - x_1^*), (\beta_0 + \beta_1 x_1^* + \beta_2 x_2^* + \beta_3'c)(x_1 - x_1^*)$$
$$, (\delta_0 + \delta_1 x_1^* + \delta_2 x_2^* + \delta_3'c)(x_2 - x_2^*), (\beta_0 + \beta_1 x_1^* + \beta_2 x_2^* + \beta_3'c)(x_2 - x_2^*), 0')$$

Thus the standard error of the direct effect in model 2:

$$\sqrt{\nabla g(\theta)' Var(\theta) \nabla g(\theta)}$$

Given the indirect effect for model 2 in Section 3 we have

$$g(\theta) = ((\phi_3 + \gamma_3 t)\delta_1 + (\phi_4 + \gamma_4 t)\beta_1)(x_1 - x_1^*) + ((\phi_3 + \gamma_3 t)\delta_2 + (\phi_4 + \gamma_4 t)\beta_2)(x_2 - x_2^*)$$
$$+ ((\phi_5 + \gamma_5 t)\delta_1 + (\phi_6 + \gamma_6 t)\beta_1)(x_1 x_1 - x_1 x_1^*) + ((\phi_5 + \gamma_5 t)\delta_2 + (\phi_6 + \gamma_6 t)\beta_2)(x_1 x_2 - x_1 x_2^*)$$
$$+ ((\phi_7 + \gamma_7 t)\delta_1 + (\phi_8 + \gamma_8 t)\beta_1)(x_2 x_1 - x_2 x_1^*) + ((\phi_7 + \gamma_7 t)\delta_2 + (\phi_8 + \gamma_8 t)\beta_2)(x_2 x_2 - x_2 x_2^*)$$

Thus we have

$$\nabla g(\theta) = (\ 0,0,0, (\phi_3 + \gamma_3 t)(x_1 - x_1^*) + (\phi_5 + \gamma_5 t)(x_1 x_1 - x_1 x_1^*) + (\phi_7 + \gamma_7 t)(x_2 x_1 - x_2 x_1^*)$$
$$, (\phi_3 + \gamma_3 t)(x_2 - x_2^*) + (\phi_5 + \gamma_5 t)(x_1 x_2 - x_1 x_2^*) + (\phi_7 + \gamma_7 t)(x_2 x_2 - x_2 x_2^*), 0', 0$$
$$, (\phi_4 + \gamma_4 t)(x_1 - x_1^*) + (\phi_6 + \gamma_6 t)(x_1 x_1 - x_1 x_1^*) + (\phi_8 + \gamma_8 t)(x_2 x_1 - x_2 x_1^*)$$
$$, (\phi_4 + \gamma_4 t)(x_2 - x_2^*) + (\phi_6 + \gamma_6 t)(x_1 x_2 - x_1 x_2^*) + (\phi_8 + \gamma_8 t)(x_2 x_2 - x_2 x_2^*)$$
$$, 0', 0, 0, 0, \delta_1(x_1 - x_1^*) + \delta_2(x_2 - x_2^*), \beta_1(x_1 - x_1^*) + \beta_2(x_2 - x_2^*)$$
$$, \delta_1(x_1 x_1 - x_1 x_1^*) + \delta_2(x_1 x_2 - x_1 x_2^*), \beta_1(x_1 x_1 - x_1 x_1^*) + \beta_2(x_1 x_2 - x_1 x_2^*)$$
$$, \delta_1(x_2 x_1 - x_2 x_1^*) + \delta_2(x_2 x_2 - x_2 x_2^*), \beta_1(x_2 x_1 - x_2 x_1^*) + \beta_2(x_2 x_2 - x_2 x_2^*)$$
$$, 0', 0, 0, 0, \delta_1 t(x_1 - x_1^*) + \delta_2 t(x_2 - x_2^*), \beta_1 t(x_1 - x_1^*) + \beta_2 t(x_2 - x_2^*)$$
$$, \delta_1 t(x_1 x_1 - x_1 x_1^*) + \delta_2 t(x_1 x_2 - x_1 x_2^*), \beta_1 t(x_1 x_1 - x_1 x_1^*) + \beta_2 t(x_1 x_2 - x_1 x_2^*)$$
$$, \delta_1 t(x_2 x_1 - x_2 x_1^*) + \delta_2 t(x_2 x_2 - x_2 x_2^*), \beta_1 t(x_2 x_1 - x_2 x_1^*) + \beta_2 t(x_2 x_2 - x_2 x_2^*), 0'$$

Thus the standard error of the indirect effect in model 2:

$$\sqrt{\nabla g(\theta)' Var(\theta) \nabla g(\theta)}$$

# 5 An Example

In this section we give an example of a longitudinal mediation analysis using latent growth curve models and the definition of the direct and indirect effects shown in section 2. The data and motivation of this example comes from Gunzler et al.[27]. Their goal was to develop an adjusted screening tool to better assess depressive symptoms in Multiple Sclerosis (MS) patients. Screening for depression in this population can be challenging due to the overlap of MS symptoms with symptoms of depression. Disentangling these relationships can be key for treatment as depression is the most frequent psychiatric diagnosis in MS patients [28].

Consider the latent growth curve model as shown in figure 3. We are interested in how MS type ($0 \rightarrow$ relapsing, $-1 \rightarrow$ progessive) affects self-reported depression screening (PHQ-9) directly and indirectly through a timed 25-foot walk. As noted in Figure 3, log timed walk (ltw) and PHQ-9 (PHQ) are measured at 6 different time points. These time points vary between subjects.

PHQ-9 is used both in screening and monitoring of depression in patients. Patients respond to a likert scale from 0 (not at all) to 3 (every day) about 9 different symptoms over the prior 2 weeks before their appointment[27]. This leads to a total score with a range from 0 to 27. The 25-foot timed walk is a quantitative test of mobility and leg function. Gunzler et al used an additional peg test to quantitatively assess arm and hand function as well as using each symptom of PHQ-9 as an outcome. Here we focus on the total PHQ-9 score as the outcome and only the log timed walk for the mediator. .

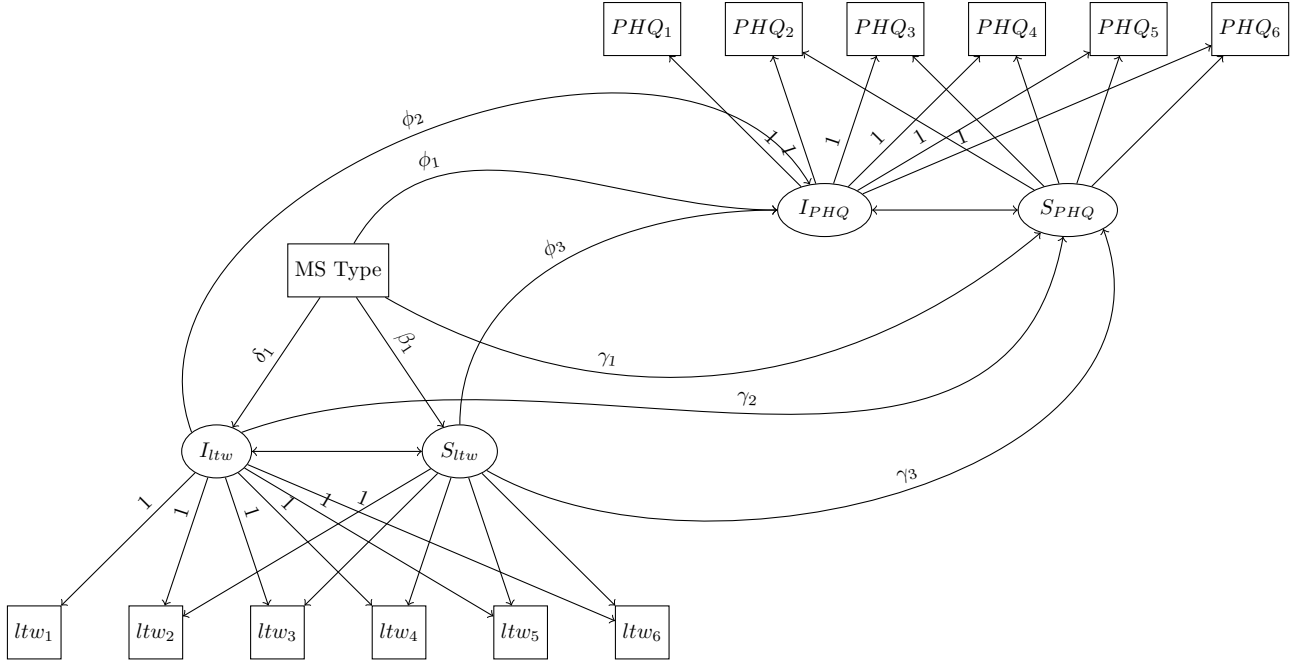

Figure 3: MS and Depression Example

Gunzler et al.[27] used data from the Knowledge Program developed at Cleveland Clinic's Neurological Institute [29] which links PHQ-9 data to its EPIC electronic health record. The Mellen Center[30] for MS manages more than 20,000 visits and 1,000 new patients every year for MS treatment. The Knowledge Program tracks illness severity and treatment efficacy over time across the Mellen Center. This data comes from a retrospective cohort containing patients with measurements of PHQ-9 and a 25-foot timed walk data available. Table 1 displays the demographic information of the 3,507 patients in the sample from 2008 - 2011. In the table the patients are split by a PHQ-9 score of $< 10$ and $\geq 10$, where 10 is a validated threshold for moderate depression[31].

For this example we fit the following mediator process

$$ltw_{it} = I_{ltw_i} + S_{ltw_i} + \varepsilon_{ltw_{it}}$$

$$I_{ltw_i} = \delta_0 + \delta_1 CCLB1_i + \nu_{I_{ltw_i}}$$

$$S_{ltw_i} = \beta_0 + \beta_1 CCLB1_i + \nu_{S_{ltw_i}}$$

and the following outcome process

$$PHQ_{it} = I_{PHQ_i} + S_{PHQ_i} + \varepsilon_{PHQ_{it}}$$

$$I_{PHQ_i} = \phi_0 + \phi_1 CCLB1_i + \phi_2 I_{ltw_i} + \phi_3 I_{Stw_i} + \nu_{I_{PHQ_i}}$$

|  | PHQ $-$ 9 < 10<br>$n = 2{,}502$ | PHQ $-$ 9 $\geq$ 10<br>$n = 1{,}005$ | P-value |
|---|---|---|---|
| PHQ-9 | $3.64 \pm 2.75$ | $15.26 \pm 4.40$ | < 0.001 |
| MSPS fatigue | $1.62 \pm 1.25$ | $3.35 \pm 1.12$ | < 0.001 |
| MSPS cognitive | $0.86 \pm 0.96$ | $2.23 \pm 1.30$ | < 0.001 |
| MSPS mobility | $1.37 \pm 1.58$ | $2.39 \pm 1.48$ | < 0.001 |
| MSPS hand function | $0.77 \pm 0.94$ | $1.79 \pm 1.27$ | < 0.001 |
| 25-Foot time walk | $7.85 \pm 10.56$ | $8.83 \pm 7.61$ | 0.002 |
| 9-hole peg test | $23.68 \pm 10.66$ | $26.82 \pm 12.48$ | < 0.001 |
| Age | $46.12 \pm 11.88$ | $44.47 \pm 11.20$ | < 0.001 |
| Baseline time since diagnosis | $11.80 \pm 10.00$ | $10.89 \pm 9.37$ | 0.016 |
| Female $n(\%)$ | 1,836(74) | 740(74) | 0.879 |
| Race, $n(\%)$ |  |  | 0.07 |
| Caucasian | 2,112 (85) | 821 (82) |  |
| African-American | 225 (9) | 114 (11) |  |
| Other | 144 (6) | 65 (7) |  |
| MS type, $n(\%)$ |  |  | 0.067 |
| Relapsing | 2,045 (84) | 787 (82) |  |
| Progressive | 383 (16) | 177 (18) |  |

Table 1: Demographics of the 3,507 Patients in Sample

$$S_{PHQ_i} = \gamma_0 + \gamma_1 CCLB1_i + \phi_2 I_{ltw_i} + \phi_3 I_{Stw_i} + \nu_{S_{ltw_i}}.$$

Where $\mathbf{E}\left[\varepsilon_{ltwit}\right] = \mathbf{E}\left[\varepsilon_{PHQ_{it}}\right] = \mathbf{E}\left[\nu_{I_{ltw_i}}\right] = \mathbf{E}\left[\nu_{S_{ltw_i}}\right] = \mathbf{E}\left[\nu_{I_{PHQ_i}}\right] = \mathbf{E}\left[\nu_{S_{PHQ_i}}\right] = 0$ and where $\varepsilon_{ltw_{it}}, \varepsilon_{PHQ_{it}}, (\nu_{I_{ltw_i}}, \nu_{S_{ltw_i}})$ and $(\nu_{I_P}$ are mutually independent. For simplicity here were no baseline covariates adjusted for in the analysis.

| Variable | Estimate | Std. Err. | $t$ | $\Pr > \lvert t \rvert$ | 95% CI | |
|---|---|---|---|---|---|---|
| $\delta_0$ | 2.361 | 0.029 | 81.106 | < 0.001 | 2.30416 | 2.41784 |
| $\delta_1$ | 0.59 | 0.031 | 19.057 | < 0.001 | 0.52924 | 0.65076 |
| $\beta_0$ | 0.066 | 0.012 | 5.695 | < 0.001 | 0.04248 | 0.08952 |
| $\beta_1$ | 0.049 | 0.012 | 4.018 | < 0.001 | 0.02548 | 0.07252 |
| $\phi_0$ | -0.825 | 0.996 | -0.828 | 0.408 | -2.77716 | 1.12716 |
| $\phi_1$ | -1.238 | 0.377 | -3.288 | 0.001 | -1.97692 | -0.49908 |
| $\phi_2$ | 3.677 | 0.499 | 7.365 | < 0.001 | 2.69896 | 4.65504 |
| $\phi_3$ | -8.897 | 4.343 | -2.049 | 0.04 | -17.4093 | -0.38472 |
| $\gamma_0$ | 1.643 | 0.463 | 3.552 | < 0.001 | 0.73552 | 2.55048 |
| $\gamma_1$ | 0.396 | 0.18 | 2.205 | 0.027 | 0.0432 | 0.7488 |
| $\gamma_2$ | -0.904 | 0.233 | -3.875 | < 0.001 | -1.36068 | -0.44732 |
| $\gamma_3$ | 8.185 | 2.096 | 3.905 | < 0.001 | 4.07684 | 12.29316 |

Table 2: Estimates from Model showin if Figure 3. Obtained using Mplus version 7.2[32]

Table 2 displays the results estimated by fitting the above model in Mplus. Recall from Section 2 that for this model without interaction the direct effect is $(\phi_1 + \gamma_1 t)(x - x^*)$ and the indirect effect is $((\phi_2 + \gamma_2 t)\delta_1 + (\phi_3 + \gamma_3 t)\beta_1)(x - x^*)$. We let $x = 0$ and $x^* = -1$ to reflect a change in MS status from relapsing to progressive such that the direct effect is

$$(\phi_1 + \gamma_1 t)(x - x^*) = -1.238 + 0.396t$$

and the indirect effect is

$$((\phi_2 + \gamma_2 t)\delta_1 + (\phi_3 + \gamma_3 t)\beta_1)(x - x^*) = ((3.677 - 0.904t)0.59 + (-8.897 + 8.185t)0.049)$$

Recall from Section 4.1 that the standard error of the direct effect is $\sqrt{\nabla g(\theta_d)' Var(\theta) \nabla g(\theta_d)}|x - x^*|$ where $\nabla g(\theta_d)' = (0, 0, 0, 0, 0, 1, 0, 0, 0, t, 0, 0)$ and the standard error for the indirect effect is $\sqrt{\nabla g(\theta_{id})' Var(\theta) \nabla g(\theta_{id})}$ where $\nabla g(\theta_{id})' = ((0, (\phi_2 + \gamma_2 * t)(x - x^*), 0, (\phi_3 + \gamma_3 * t)(x - x^*), 0, 0, (\delta_1, \beta_1)(x - x^*), 0, 0, \delta_1 * t(x - x^*), \beta_1 * t(x - x^*))$, where $Var(\theta)$ is estimated by Mplus.

Table 3 displays the direct and indirect effects at various time points as well as a 95% confidence interval at each time point and p-value of the effect at that particular time point. We see that initially direct effect is negative however it becomes statistically insignificant sometime between 1 and 2 years. However the indirect effect is positive with decreasing effect size yet remains statistically significant throughout the duration of this study.

| Time (years) | Direct Effect | 95%CI Lower | 95%CI Upper | p-value | Indirect Effect | 95%CI Lower | 95%CI Upper | |
|---|---|---|---|---|---|---|---|---|
| 0 | -1.238 | -1.97659 | -0.49942 | 0.001 | 1.733477 | 1.270811 | 2.196144 | < 0.001 |
| 1 | -0.842 | -1.50667 | -0.17733 | 0.013 | 1.601182 | 1.245434 | 1.95693 | < 0.001 |
| 2 | -0.446 | -1.21115 | 0.319154 | 0.253 | 1.468887 | 1.065199 | 1.872575 | < 0.001 |
| 3 | -0.05 | -1.0382 | 0.938198 | 0.921 | 1.336592 | 0.768012 | 1.905172 | < 0.001 |

Table 3: Direct and Indirect Effects of Model in Figure 3

MacKinnon[18] defines the direct effect as $\gamma_1 = 0.396$ 95% CI (0.043, 0.749) and an indirect effect of $\beta_1 \gamma_3 = (0.049)(8.185) = 0.401$ 95% CI (0.122, 0.680), where the standard error of the indirect effect is $\sqrt{\beta_1^2 \sigma_{\gamma_3}^2 + \gamma_3^2 \sigma_{\beta_1}^2} = 0.142$. This makes the further assumption that the direct and indirect effect remain constant throughout time as opposed to the methods in this paper which allow for the direct and indirect effect to change with time.

# 6 Discussion

This paper mathematically defines the direct and indirect effects of longitudinal mediation with latent growth curve models using counterfactuals. We build upon the models considered by MacKinnon[17, 18] but allowed for the presence of treatment/exposure-mediator interaction. We then considered the assumptions needed for identifiability of these direct and indirect effects. Those assumptions are:

1. No unmeasured confounding of the exposure-outcome relationship

2. No unmeasured confounding of the mediator-outcome relationship

3. No unmeasured confounding of the exposure-mediator relationship

4. No mediator-outcome confounders which are affected by the exposure

We mathematically define these effects using the above assumptions first with a model in which the treatment/exposure is binary, followed by a model in which the treatment/exposure itself changes with time. We find that latent growth mediation models in current literature allow for the intercept of the outcome to be a cause for the slope of the mediator. This violates assumption 4 since the intercept of the outcome would become a mediator-outcome confounder which itself was affected by exposure. We also find that with models currently in the literature it is assumed that the direct and indirect effects remain constant while these methods allow them to vary with time. With the direct and indirect effects defined we consider the delta method for estimating the standard error of those effects.

We then gave an example using model 1 and found the direct and indirect effects along with their 95% confidence intervals. We compared this to the direct and indirect effect estimates obtained using MacKinnon's method.

# Acknowledgments

Research reported in this publication was supported by the National Institutes of Health under award numbers T32NS048005 and R01ES017876. The content is solely the responsibility of the authors and does not necessarily represent the official views of the National Institutes of Health.

31. K. Kroenke and R. L. Spitzer, "The phq-9: a new depression diagnostic and severity measure," *Psychiatr Ann*, vol. 32, no. 9, pp. 1–7, 2002.

32. B. O. Muthén and L. K. Muthén, *Mplus (Version 7.2) [Computer software]*, 1998–2014).
24